\documentclass[10pt,final,doublecolumn]{IEEEtran}
\usepackage{amsmath}
\usepackage{latexsym}
\usepackage{graphicx}
\usepackage{bbding}
\usepackage{indentfirst}
\usepackage{algorithm,algorithmic}
\usepackage{setspace}
\usepackage{float}
\usepackage{epstopdf}
\usepackage{amssymb}
\usepackage{amsfonts}
\usepackage{enumerate}
\usepackage{multicol}
\usepackage{color}
\usepackage{bm}
\usepackage{amssymb}
\usepackage{stfloats}
\usepackage{epstopdf}
\usepackage{threeparttable}
\usepackage{epstopdf}
\usepackage{threeparttable}
\usepackage{subfigure}
\usepackage{makecell}
\usepackage{cite}
\usepackage{verbatim}
\IEEEoverridecommandlockouts
\allowdisplaybreaks[4]
\begin{document}
	\title{Capacity Maximization for Base Station with Hybrid Fixed and Movable Antennas}	\author{Xiaoming Shi,~Xiaodan Shao,~\IEEEmembership{Member,~IEEE,}~and~Rui Zhang,~\IEEEmembership{Fellow,~IEEE,}
		% <-this % stops a space
		\thanks{X. Shi is with Shenzhen Research
			Institute of Big Data, School of Science and Engineering, The Chinese University of Hong Kong, Shenzhen,
			Guangdong 518172, China
			(e-mail: xiaomingshi@link.cuhk.edu.cn).}% <-this % stops a space
		\thanks{X. Shao is with School of Science and Engineering, The Chinese University of Hong Kong, Shenzhen, Guangdong 518172, China. She is also with the Institute for Digital Communications, Friedrich-Alexander-University Erlangen-Nuremberg, 91054 Erlangen, Germany (e-mail: xiaodan.shao@fau.de). }
		\thanks{R. Zhang is with School of Science and Engineering, Shenzhen Research
			Institute of Big Data, The Chinese University of Hong Kong, Shenzhen,
			Guangdong 518172, China (e-mail: rzhang@cuhk.edu.cn). He is also with the Department of Electrical and
			Computer Engineering, National University of Singapore, Singapore 117583
			(e-mail: elezhang@nus.edu.sg). 
			\emph{(Corresponding author: Xiaodan Shao and Rui Zhang)}}
		}
	\maketitle
\begin{abstract}
Six-dimensional movable antenna (6DMA) is an effective solution for enhancing wireless network capacity through the adjustment of both 3D positions and 3D rotations of distributed antennas/antenna surfaces. 
Although freely positioning/rotating 6DMA surfaces offers the greatest flexibility and thus highest capacity improvement, its implementation may be challenging in practice due to the drastic architecture change required for existing base stations (BSs), which predominantly adopt fixed-position antenna (FPA) arrays (e.g., sector antenna arrays). 
Thus, we introduce in this letter a new BS architecture called hybrid fixed and movable antennas (HFMA), which consists of both conventional FPA arrays and position/rotation-adjustable 6DMA surfaces.
For ease of implementation, we consider that all 6DMA surfaces can rotate  along a circular track above the FPA arrays. 
We aim to maximize the network capacity via optimizing the rotation angles of all 6DMA surfaces based on the users' spatial distribution.
Since this problem is combinatorial and its optimal solution requires prohibitively high computational complexity via exhaustive search, we propose an alternative adaptive Markov Chain Monte Carlo based method to solve it more efficiently. 
Finally, we present simulation results that show significant performance gains achieved by our proposed design over various benchmark schemes.
\end{abstract}
	

\vspace{-15pt}
\section{Introduction}
The future sixth-generation (6G) wireless networks are envisioned as an enabler for various emerging applications, such as extended reality, intelligent transportation, and massive Internet of Things (IoT).
These applications are expected to lead to 30 billion or more IoT devices by 2030, requiring increasingly higher network capacity and transmission reliability \cite{9509294}.
To achieve this challenging goal, advanced multiple-input multiple-output (MIMO) communication  techniques, such as cell-free massive MIMO \cite{7827017}, extremely large-scale MIMO \cite{8732419}, and intelligent reflecting surface (IRS)-aided MIMO \cite{qingqing}, have been proposed to improve the performance of wireless communication systems by leveraging their spatial degrees of freedom (DoFs).
However, existing MIMO systems mostly adopt fixed-position antennas (FPAs) at the base stations (BSs) and user terminals (UTs), which cannot fully exploit the spatial variation of wireless channels at the transceivers. 
Thus, movable antenna (MA) system \cite{10318061,10286328,10414081}, also known as fluid antenna system (FAS) \cite{9650760,zhu2024historical}, was introduced to enable the antenna movement over a confined region at the BS/UT for achieving more favorable channel conditions to improve the communication performance. However, the existing works on MAs/FAS have mainly considered the antenna position adjustment on a given 2D surface or a given line, thus not fully exploiting the spatial DoFs in a given 3D space. 

Recently, to enable the full flexibility in antenna deployment at the BS, six-dimensional movable antenna (6DMA) system has been proposed as a new and effective solution for improving wireless network capacity \cite{shao20246d,6dma_dis}.
Equipped with distributed 6DMA surfaces which can be jointly controlled in 3D positions as well as 3D rotations to cater to the users' spatial distribution, the 6DMA-empowered BS can maximally utilize the antennas' directionality, array gain and spatial multiplexing gain to significantly enhance the wireless network capacity. 
In practice, the positions and/or rotations of 6DMA surfaces can be adjusted continuously \cite{shao20246d} or in discrete levels \cite{6dma_dis} (depending on the surface movement mechanism), and optimized with or without the prior knowledge of users' spatial distribution in the network \cite{shao20246d,6dma_dis}.

However, despite the high performance gains achievable with 6DMA-BS, it faces challenges in practical implementation as the antenna architecture of BSs requires a fundamental reform from the existing FPA arrays (e.g., sector antenna arrays) to the fully-adjustable 6DMA surfaces, which may result in a significant increase of network infrastructure cost.   
To alleviate the high implementation cost of 6DMA, we propose in this letter a more cost-effective BS architecture called hybrid fixed and movable antennas (HFMA), which consists of both FPA arrays and 6DMA surfaces.
In particular, for ease of implementation, we consider that all 6DMA surfaces can rotate along a circular track above the FPA arrays.
With the new HFMA at the BS, the users in hotspot areas of the cell can be served by the 6DMA surfaces flexibly by properly adjusting their rotation angles, while the other (regular) users in the remaining areas of the cell can be served by conventional FPAs, thus reducing the overall antenna deployment cost at the BS as compared to the 6DMA-BS.
To maximize the network capacity, we formulate an optimization problem by jointly designing the rotation angles of all 6DMA surfaces based on the {\it a priori} known users' spatial distribution.
However, this problem is combinatorial and thus requires prohibitively high complexity to solve optimally via exhaustive search. 
To reduce the computational complexity, we propose an alternative method to solve it more efficiently based on the adaptive Markov Chain Monte Carlo (AMCMC) technique \cite{liu2001monte}.
Simulation results show that the proposed HFMA-BS scheme and AMCMC-based solution  outperform various benchmark schemes in terms of the network capacity by optimally  rotating/allocating 6DMA surfaces at the BS to match the (non-uniform) users' spatial distribution.

\textit{Notations}: Boldface lower-case letter denotes vector, $(\cdot)^H$ and $(\cdot)^T$ denote conjugate transpose and transpose, respectively, $\mathbb{E}[\cdot]$ denotes the expected value of random variable, ${\bf 0}_{N}$ and ${\bf 1}_{N}$ denote the $N \times 1$ vector with all zero elements and all one elements, respectively, ${\bf I}_{N}$ denotes the $N \times N$ identity matrix, $[ {\bf a}]_i$ denotes the $i$-th element of vector $\bf a$, $supp(\mathbf{v})$ denotes the set of indices of the nonzero entries in vector $\mathbf{v}$,  $\lfloor \cdot \rfloor$ denotes the floor operator, $\otimes$ denotes the Kronecker product, and $U[a,b]$ denotes the uniform distribution within real-number interval $[a,b]$.

\vspace{-3 pt}
\section{System Model and Problem Formulation}
\begin{figure*}[!t]	
	\centering
	\subfigure[HFMA-BS]{\includegraphics[width=2.5in]{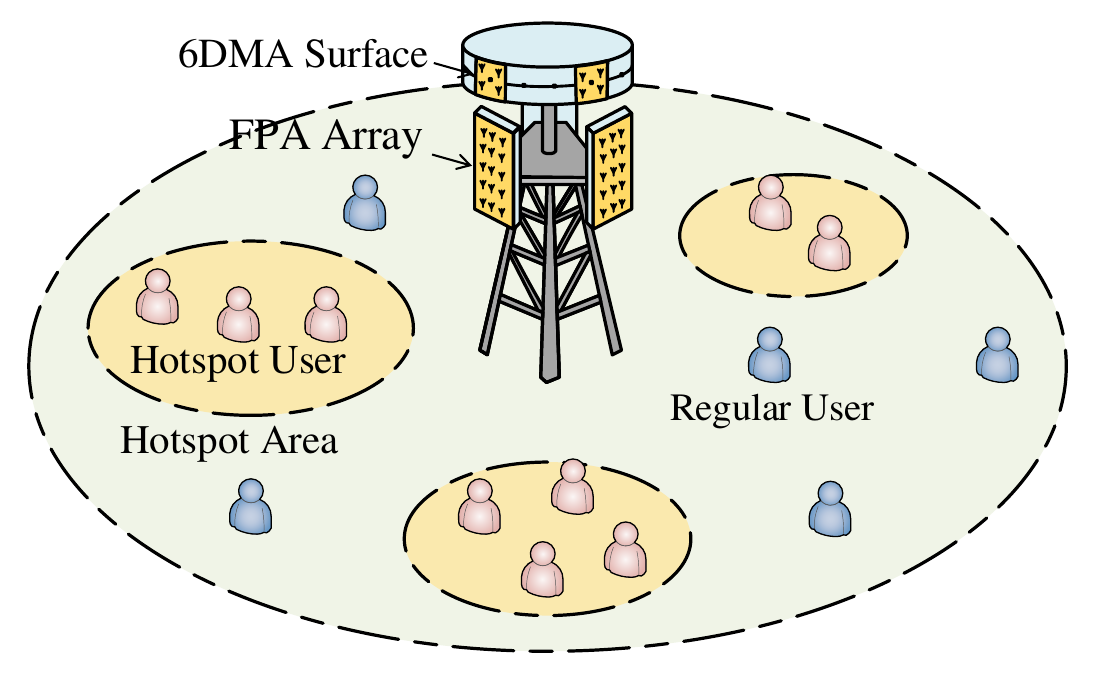}}
	\hspace{0.1 in}
	\subfigure[FPA array]{\includegraphics[width=1.3in]{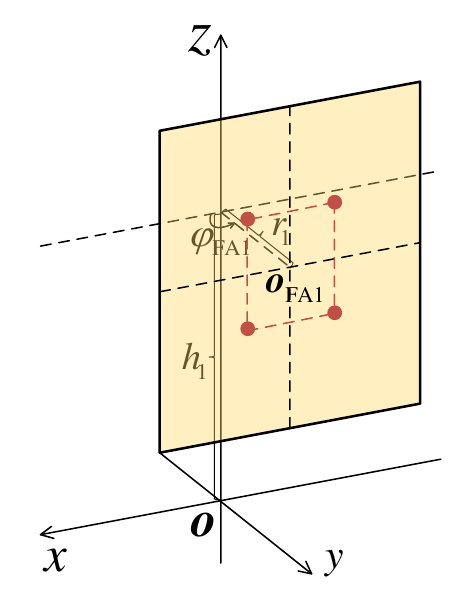}}
	\hspace{0.1 in}
	\subfigure[6DMA surface]{\includegraphics[width=2.0in]{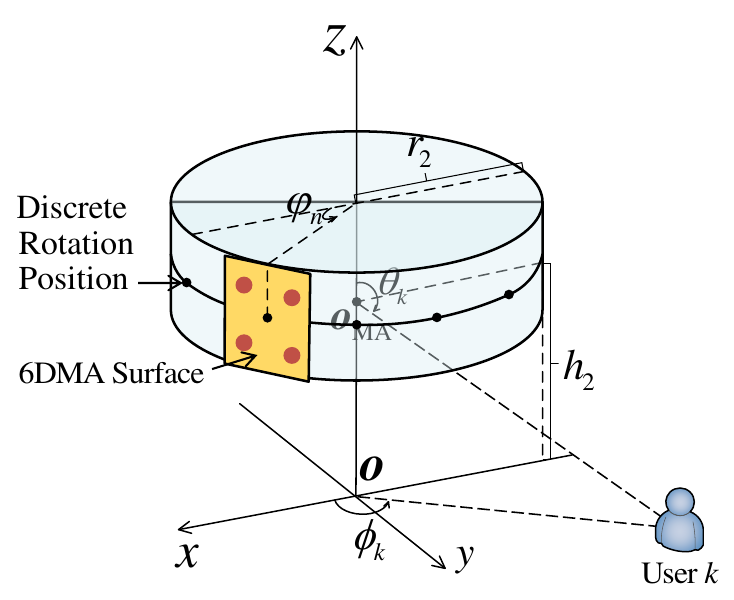}}
	\vspace{-4 pt}
	\caption{Illustration of the HFMA-BS and the geometries of FPA array and 6DMA surface.}\label{scenario}
\end{figure*}

\subsection{HFMA-BS}
We consider the uplink transmission from $K$ ground users, denoted by the set $\mathcal{K}=\{1,2,\cdots,K\}$, to a cellular BS which is assumed to be equipped with the HFMA.
Each user is equipped with a single isotropic antenna.
As shown in Fig. \ref{scenario}(a), the antennas at the HFMA-BS consist of two parts, namely, one part with three FPA arrays (assumed to be conventional sector antenna arrays), and the other part  with $N_{\rm MA}$ 6DMA surfaces, denoted by the set ${\mathcal{N}}_{\rm MA}=\{1,2,\cdots,N_{\rm MA}\}$.
The FPA arrays are assumed to be uniform planar arrays (UPAs), each comprising $M_{\rm FA}=M_{{\rm FA},h}\times M_{{\rm FA},v}$ antennas, where $M_{{\rm FA},h}$ and $M_{{\rm FA},v}$ denote the number of antennas in the horizontal direction and the vertical direction, respectively.
Each FPA array has a fixed downtilt of zero degree, with its center positioned at a height of $h_1$ and a horizontal distance $r_1$ from the ground center of the BS, denoted by $\bf o$ (see Fig. \ref{scenario}(b)).
For ease of description, we define the fixed rotation angle of each FPA array as the azimuth angle measured from the $x$-axis in the Cartesian coordinate system (CCS) $o$-$xyz$ to the FPA array's center, denoted by $\varphi_{{\rm FA},f}\in [0,2\pi)$, $f=1,2,3$.

The 6DMA surfaces are also modeled as UPAs, each comprising $M_{\rm MA}=M_{{\rm MA},h}\times M_{{\rm MA},v}$ antennas, where $M_{{\rm MA},h}$ and $M_{{\rm MA},v}$ denote the number of antennas in the horizontal direction and the vertical direction, respectively.
We define the rotation angle of each 6DMA surface as the azimuth angle measured from the $x$-axis to the the 6DMA surface's center, denoted by $\varphi_n\in [0,2\pi), n\in {\mathcal{N}}_{\rm MA}$ (see Fig. \ref{scenario}(c)).
The 6DMA surfaces are connected to a central processing unit (CPU) at the BS via flexible cables \cite{shao20246d,6dma_dis} and can move along a circular track that is above the FPA arrays and  parallel to the ground.
The circle's center is at the Cartesian coordinate ${\bf o}_{\text {MA}}=\left [0,0,h_2\right ]^{ T}$ and its radius is $r_2$.

Denote the set of rotation angles for all 6DMA surfaces as ${ {\boldsymbol{ \varphi} }}=\left[ {\varphi}_1,{\varphi}_2,\cdots,{\varphi}_{N_{\rm MA}} \right]^{T}$ and the total number of antennas at the BS as $M=M_{\rm MA}N_{\rm MA}+3M_{\rm FA}$.
Then, the channel vector from user $k$, $k\in\mathcal{K}$, to the BS can be expressed as ${\bf h}_k({  {\boldsymbol{ \varphi} }})\in \mathbb{C}^{M\times 1}$, which is a function of ${  {\boldsymbol{ \varphi} }}$ in general.
The received signal vector at the BS is thus given by
	\vspace{-4pt}
\begin{equation}
	\vspace{0 pt}
	{\bf y}({  {\boldsymbol{ \varphi} }}) = \sum\limits_{k = 1}^K {{\bf h}_k({  {\boldsymbol{ \varphi} }}){s_k} + {\bf n}},
	\vspace{-4pt}
\end{equation}
where $s_k$ is the transmitted signal from user $k$ with $\mathbb{E}\{{\left| {{s_k}} \right|^2}\}=P_0$ and $P_0$ denoting the transmit power; ${\bf n}\sim \mathcal{C}\mathcal{N}({\bf 0}_{M},\sigma ^2{\bf{I}}_{M})$ denotes the independent and identically distributed (i.i.d.) complex additive white Gaussian noise (AWGN) vector with zero mean and average power $\sigma ^2$.

\vspace{-4 pt}
\subsection{Channel Model}
\vspace{0 pt}
We set the BS's reference position as the center of the circular track, ${\bf o}_{\text {MA}}$, and denote $\phi_k$ and $\theta_k$ as the azimuth and elevation angles of the signal from user $k$ arriving at the reference position, respectively.
For user $k$ located at the Cartesian coordinate $[x_k,y_k,0]^{ T}$, $\theta_k$ and $\phi_k$ can be  respectively expressed as $\theta_k = \arctan ( {{h_2}/{{\sqrt {x_k^2 + y_k^2} }}})+\frac{\pi}{2}$ and
\begin{equation}
	{\phi _k} = \left\{ {\begin{array}{*{20}{l}}
			{\arctan ( {\frac{{{y_k}}}{{{x_k}}}} ),}&{x_k > 0,y_k > 0,}\\
			{\arctan ( {\frac{{{y_k}}}{{{x_k}}}} ) + 2\pi ,}&{x_k > 0,y_k < 0,}\\
			{\arctan ( {\frac{{{y_k}}}{{{x_k}}}} )+ \pi, }&{x_k < 0,}
	\end{array}} \right.
\end{equation}
where $\theta_k\in \left[\frac{\pi}{2},\pi\right]$ and $\phi_k\in \left[0,2\pi\right]$.
Given $\theta_k$, $\phi_k$ and fixed $\varphi_{{\rm FA},f}$, the overall array response vector ${\bf \tilde f}_k({  {\boldsymbol{ \varphi} }})\in \mathbb{C}^{M\times 1}$ at the BS can be defined as
\vspace{-2 pt}
\begin{equation}\vspace{-2 pt}
	{\bf \tilde f}_k({  {\boldsymbol{ \varphi} }})=\big[ {\bf \tilde f}_{{\rm MA},k}^{T}({  {\boldsymbol{ \varphi} }}),{\bf \tilde f}_{{\rm FA},k}^{T} \big]^{T},
\end{equation}
with
\vspace{-4 pt}
\begin{equation}\vspace{-2 pt}
	{\bf \tilde f}_{{\rm MA},k}({  {\boldsymbol{ \varphi} }})=\left[ {\bf f}_{{\rm MA},k}^{T}({  \varphi }_1),\cdots,{\bf f}_{{\rm MA},k}^{T}({  \varphi}_{N_{\rm MA}}) \right]^{T},
\end{equation}
and
\vspace{-2 pt}
\begin{equation}\vspace{-2 pt}
	{\bf \tilde f}_{{\rm FA},k}=\left[ {\bf f}_{{\rm FA},k}^{T}(\varphi_{\rm FA,1}),{\bf f}_{{\rm FA},k}^{T}(\varphi_{\rm FA,2}),{\bf f}_{{\rm FA},k}^{T}(\varphi_{\rm FA,3}) \right]^{T},
\end{equation}
where ${\bf \tilde f}_{{\rm MA},k}({  {\boldsymbol{ \varphi} }})$ and ${\bf \tilde f}_{{\rm FA},k}$ denote the array response vector of all 6DMA surfaces and all FPA arrays, respectively.
We assume that the antennas on each FPA/6DMA surface are symmetrically arranged around the array center with an inter-antenna distance of $\lambda_c/2$, where $\lambda_c$ denotes the carrier wavelength.
Given $\varphi_n$, the array response vector of the $n$-th 6DMA surface is given by
%the position of the antenna in the $i$-th row and $j$-th column on the $n$-th 6DMA surface can be expressed as ${\bf t}^n_{i,j}(\varphi_n)=[r_2\cos {\varphi^n _{i,j}},r_2\sin {\varphi^n_{i,j}},z^n_{i,j}+h_2]^{ T}$, where $\varphi _{i,j}^n = \big( {j - \frac{{{M_{{\rm MA},h}} + 1}}{2}} \big)\frac{{\lambda_c }}{2r_2} + {\varphi _n}, i=1,...,M_{{\rm MA},v}, j=1,...,M_{{\rm MA},h}$ denotes the rotation angle of the $(i,j)$-th antenna on the $n$-th 6DMA surface, and $z_{i,j}^n = \big( {\frac{{{M_{{\rm MA},v}} + 1}}{2} - i} \big)\frac{{{\lambda _c}}}{2}$ denotes the height difference between the $(i,j)$-th antenna and the center of the $n$-th 6DMA surface.
\begin{equation}
	{\bf f}_{{\rm MA},k}(\varphi_n)=\sqrt{G_k(\varphi_n)}{e^{j\rho_{1,k}(\varphi_n)  }}{\bf a}_{{\rm MA},h}({\varphi}_{n}) \otimes {\bf a}_{{\rm MA},v},
\end{equation}
with
\begin{equation}
		{\bf a}_{{\rm MA},h}({\varphi}_{n})=\left[ {\begin{array}{*{20}{c}}
				{{e^{j\pi \left( {\frac{{{M_{{\rm MA},h}} + 1}}{2} - 1} \right)\sin \left( {{\varphi_n} - {\phi _k}} \right)\sin \left( {{\theta _k}} \right)}}}\\
				\vdots \\
				{{e^{j\pi \left( {\frac{{{{M_{{\rm MA},h}}} + 1}}{2} - {M_{{\rm MA},h}}} \right)\sin \left( {{\varphi_n} - {\phi _k}} \right)\sin \left( {{\theta _k}} \right)}} }
		\end{array}} \right],
\end{equation}
and
\begin{equation}
	{\bf a}_{{\rm MA},v}=
	\left[ {\begin{array}{*{20}{c}}
			{{e^{j\pi \left( {\frac{{{{M_{{\rm MA},v}}} + 1}}{2} - 1} \right)\cos \left( {{\theta _k}} \right)}}}\\
			\vdots \\
			{{e^{j\pi \left( {\frac{{{{M_{{\rm MA},v}}} + 1}}{2} - {M_{{\rm MA},v}}} \right)\cos \left( {{\theta _k}} \right)}}}
	\end{array}} \right],
\end{equation}
%	\begin{equation}
%		{\bf f}_{{\rm MA},k}(\varphi_n)={\bf a}_{{\rm MA},h}({\varphi}_{n}) \otimes {\bf a}_{{\rm MA},v},
%	\end{equation}
%	where
%	\begin{equation}
%		\begin{aligned}
%			&{\bf a}_{{\rm MA},h}({\varphi}_{n})=\\
%			&\left[ {\begin{array}{*{20}{c}}
%					{\sqrt{G_k(\varphi_{i,1}^n)}e^{j\frac{{2\pi }}{\lambda_c }r_2\sin(\theta_k)\cos({\phi_k}-{\varphi_{i,1}^n})}}\\
%					{\vdots}\\
%					{\sqrt{G_k(\varphi_{i,M_{{\rm MA},h}}^n)}e^{j\frac{{2\pi }}{\lambda_c }r_2\sin(\theta_k)\cos({\phi_k}-{\varphi^n_{i,M_{{\rm MA},h}}})}}
%			\end{array}} \right],
%		\end{aligned}
%	\end{equation}
%	\begin{equation}
%		{\bf a}_{{\rm MA},v}=\left [e^{j\frac{{2\pi }}{\lambda_c }z_{1,j}^n\cos(\theta_k)} ,\cdots,e^{j\frac{{2\pi }}{\lambda_c }z_{M_{{\rm MA},v},j}^n\cos(\theta_k)} \right ]^{ T}.
%	\end{equation}
where $\rho_{1,k}(\varphi_n)=\frac{{2\pi }}{{{\lambda _c}}}r_2\cos(\varphi_n-\phi_k)\sin(\theta_k)$ denotes the phase difference between the 6DMA surface's center and the BS's reference position, and $G_k(\varphi_n)$ denotes the antenna gain of each antenna with rotation angle $\varphi_n$ in the scale of dBi, defined as\cite{8638796}{\footnote{To focus on investigating the effect of the rotations of 6DMA surfaces on the network capacity, we consider the antenna's directivity in the horizontal direction only, while each antenna is assumed to be  omnidirectional in the vertical direction.}}
\begin{equation}
	G_k(\varphi_n)|{\rm {dBi}}=G_{m}-{\min \Big\{{12\Big( {\frac{{\phi_k-\varphi_n}}{{{\phi _{3{\rm{dB}}}}}}} \Big)^2},{G_s} \Big\}},
\end{equation}
where $k\in\mathcal{K}$, $G_m$ is the maximum antenna gain, ${G_s}$ is the side lobe level in the horizontal plane of the BS antenna pattern, and ${{\phi _{3{\rm{dB}}}}}$ is the half-power beamwidth in the horizontal plane.

Similarly, for each FPA array with a fixed rotation angle $\varphi_{{\rm FA},f}$, its array response vector is given by
\begin{equation}
		{\bf f}_{{\rm FA},k}(\varphi_{{\rm FA},f})\!=\!\sqrt{\!G_k(\varphi_{{\rm FA},f})}{e^{j\rho_{2,k}(\varphi_{{\rm FA},f}\!)  }}{\bf a}_{{\rm FA},h}(\varphi_{{\rm FA},f}\!) \otimes {\bf a}_{{\rm FA},v},
\end{equation}
with ${\bf a}_{{\rm FA},h}(\varphi_{{\rm FA},f})={\bf a}_{{\rm MA},h}(\varphi_{{\rm FA},f})|^{{M_{{\rm MA},h}} \to {M_{{\rm FA},h}}}$ and ${\bf a}_{{\rm FA},v} = {\bf a}_{{\rm MA},v}|^{{M_{{\rm MA},v}} \to {M_{{\rm FA},v}}}$, where $a\to b$ means replacing $a$ with $b$, and $\rho_{2,k}(\varphi_{{\rm FA},f})=\frac{{2\pi }}{{{\lambda _c}}}\big(r_1\cos(\phi_k-\varphi_{{\rm FA},f})\sin(\theta_k)-(h_2-h_1)\cos(\theta_k)\big)$ denotes the phase difference between the FPA array's center and the BS's reference position.

\begin{comment}
\begin{equation}
	\begin{aligned}
		&{\bf a}_{{\rm FA},h}(\varphi_{{\rm FA},f})=\\
		&\left[ {\begin{array}{*{20}{c}}
				{{e^{j\pi \left( {\frac{{{M_{{\rm FA},h}} + 1}}{2} - 1} \right)\sin \left( {{\varphi_{{\rm FA},f}} - {\phi _k}} \right)\sin \left( {{\theta _k}} \right)}}}\\
				\vdots \\
				{{e^{j\pi \left( {\frac{{{{M_{{\rm FA},h}}} + 1}}{2} - {M_{{\rm FA},h}}} \right)\sin \left( {{\varphi_{{\rm FA},f}} - {\phi _k}} \right)\sin \left( {{\theta _k}} \right)}} }
		\end{array}} \right],
	\end{aligned}
\end{equation}
and
\begin{equation}
	{\bf a}_{{\rm FA},v}({\varphi}_{n})=
	\left[ {\begin{array}{*{20}{c}}
			{{e^{j\pi \left( {\frac{{{{M_{{\rm FA},v}}} + 1}}{2} - 1} \right)\cos \left( {{\theta _k}} \right)}}}\\
			\vdots \\
			{{e^{j\pi \left( {\frac{{{{M_{{\rm FA},v}}} + 1}}{2} - {M_{{\rm FA},v}}} \right)\cos \left( {{\theta _k}} \right)}}}
	\end{array}} \right],
\end{equation}
\end{comment}
In this letter, for simplicity, we assume the line-of-sight (LoS) channel between each user and the BS.
As a result, the channel vector from user $k$ to the BS is given by
\begin{equation}
	{\bf h}_k({  {\boldsymbol{ \varphi} }})=\sqrt{\beta_k}e^{-j\frac{2\pi d_k}{\lambda_c}}{\bf \tilde f}_k({  {\boldsymbol{ \varphi} }}),
\end{equation}
where $\beta_k=\beta_0/d_k^2$ denotes the large-scale channel power gain between the BS and user $k$, with $\beta_0$ representing the power gain at the reference distance of $1$ meter (m), and ${d_k}$ representing the distance from user $k$ to the BS's reference position.

%From the above discussion, it can be seen that for a given user $k$, ${\bf f}_{{\rm FA},k}(\varphi_{\rm FA})$ remains constant as  the rotation angle $\varphi_{\rm FA}$ is fixed, while ${\bf f}_{{\rm MA},k}(\varphi_n)$ can be altered by adjusting the positions of the 6DMA surfaces.
%Therefore, the channel can be reconfigured, enabling signal enhancement at the BS.

\subsection{User Distribution}
In this letter, we focus on a single-cell wireless network where two types of users are present, namely regular users and hotspot users (see Fig. \ref{scenario}(a)).
We use tools from stochastic geometry to model the spatial distributions of these users.
Specifically, we model the spatial distribution of regular users over the entire cell area, denoted by $\mathcal{A}_0$, using a homogeneous Poisson point process (HPPP) with density $\mu_0$ \cite{8368129}.
In addition, we assume that there are $W$ hotspot areas in the cell, denoted by $\mathcal{A}_1,\cdots,\mathcal{A}_W$, and the spatial distribution of the users in each hotspot area is modeled as an independent HPPP with density $\mu_{w}$,  $w=1,\cdots,W$.

\subsection{Problem Formulation}
Assuming perfect channel state information (CSI) at the BS, optimal Gaussian signaling, and multiuser joint decoding, the average network capacity for the MIMO multiple-access channel (MAC) is given by \cite{1237143}
\begin{equation}\label{capacity}
	C({  {\boldsymbol{ \varphi} }})\!=\!\mathbb{E}_{{\bf H}}\bigg[\log_2 \det\! \bigg( {{{\bf I}_{M}} + \mathop \sum \limits_{k = 1}^K \frac{{{P_0}}}{{{\sigma ^2}}}{{\bf h}_k}({\boldsymbol{ \varphi} }){\bf h}_k({\boldsymbol{ \varphi} })^{\rm H}} \bigg)\bigg],
\end{equation}
in bits per second per Hertz (bps/Hz), where ${{\bf H}(\boldsymbol{ \varphi})}=[{{\bf h}_1}({  {\boldsymbol{ \varphi} }}),\cdots,{{\bf h}_K}({  {\boldsymbol{ \varphi} }})]$.
Note that different from the MIMO channel with traditional FPAs, the capacity of the HFMA-based MAC, as shown in (\ref{capacity}), depends on the rotation angles in ${ {\boldsymbol{ \varphi} }}$, which influence all the user channels, ${{\bf h}_k}({  {\boldsymbol{ \varphi} }})$, $k\in\mathcal{K}$.

Since deriving the expectation in (\ref{capacity}) analytically is challenging, we resort to employing the Monte Carlo method to approximate $C({ {\boldsymbol{ \varphi} }})$. Specifically, we generate $\Upsilon$ independent realizations of the number of users as well as their locations based on the given users' spatial distribution.
Then, the average network capacity in (\ref{capacity}) can be approximated as
\begin{equation}\label{C}
	{\hat{C}}({  {\boldsymbol{ \varphi} }})\!=\!\dfrac{1}{\Upsilon}\sum\limits_{\upsilon=1}^{\Upsilon}\log_2 \det\! \bigg( {{{\bf I}_{M}} \!+ \! { \mathop \sum \limits_{k = 1}^{K_{\upsilon}} \frac{{{P_0}}}{{{\sigma ^2}}}{{\bf h}_{k,\upsilon}}({  {\boldsymbol{ \varphi} }}){\bf h}_{k,\upsilon}({  {\boldsymbol{ \varphi} }})^{\rm H}} } \bigg),
\end{equation}
where $K_{{\upsilon}}$ is the number of users in the $\upsilon$-th realization and ${{\bf h}_{k,\upsilon}}({  {\boldsymbol{ \varphi} }}), k=1,\cdots,K_\upsilon, \upsilon=1,\cdots,\Upsilon$, is the channel vector of user $k$ in the $\upsilon$-th realization.

For ease of practical implementation, we consider that there are $L$ discrete positions equally spaced along the circular track, denoted by the set $\mathcal{L}=\{1,2,\cdots,L\}$, which can be selected for moving the $N_{\rm MA}$ 6DMA surfaces, with $L>N_{\rm MA}$.
We denote their corresponding rotation angles with respect to the center ${\bf o}_{\text {MA}}$ in the set ${\Phi}=\{{\tilde \varphi}_1,{\tilde \varphi}_2,\cdots,{\tilde \varphi}_{L}\}$, where ${\tilde \varphi}_{l}=\frac{(2{l}-1)\pi}{L},~l\in\mathcal{L}$.
In other words, the set of feasible rotation angles for the $n$-th 6DMA surface, $n\in\mathcal{N}_{\rm MA}$, is given by $\Phi$, i.e., $\varphi_n\in \Phi$. 
To avoid overlapping among 6DMA surfaces at different positions/rotations, we assume that $L\leq \lfloor \frac{2\pi r_2}{D_{\rm min}} \rfloor$, where $D_{\rm min}$ is the minimum distance required between the centers of any two 6DMA surfaces.
For clarity, we define an indicator vector for the selected rotation angles as ${\boldsymbol{\epsilon}}\in \mathbb{C}^{L\times 1}$, where $[{\boldsymbol{\epsilon}} ]_l\in\{0,1\}$ and $\sum_{l=1}^L[{\boldsymbol{\epsilon}}]_l=N_{\rm MA}$.
If 	$[{\boldsymbol{\epsilon}} ]_l=1,l\in\mathcal{L}$, it indicates that ${\tilde \varphi}_{l}$ is selected for deploying a 6DMA surface; otherwise, $[{\boldsymbol{\epsilon}} ]_l=0$.

We aim to maximize the average network capacity by determining the optimal set of rotation angles for all 6DMA surfaces.
Accordingly, the optimization problem is formulated as
\vspace{-2pt}
\begin{equation}\label{Problem}
	{\boldsymbol{\epsilon}}^*=\arg \mathop {\max }\limits_{{\boldsymbol{\epsilon}} \in {\mathcal{Z}} } ~ {\hat{C}}({\boldsymbol{ \varphi} }_{{\boldsymbol{\epsilon}}}),
	\vspace{-1 pt}
\end{equation}
where ${\boldsymbol{\epsilon}}^*$ denotes the optimal indicator vector for the selected rotation angles to maximize the capacity, ${{\mathcal{Z}}}$ denotes the set of all different indicator vectors with the cardinality of ${Q}=\binom{L}{N_{\rm MA}}$, and ${\boldsymbol{ \varphi} }_{{\boldsymbol{\epsilon}}}=[\tilde {\varphi}_{i_{1}}, \tilde {\varphi}_{i_{2}},\cdots,\tilde {\varphi}_{i_{N_{\rm MA}}}]$, with $i_{n}$, $n\in {\mathcal{N}}_{\rm MA}$, denoting the support of ${\boldsymbol{\epsilon}} $, i.e., $i_{n}\in supp({\boldsymbol{\epsilon}} )$.
Note that problem (\ref{Problem}) is a combinatorial optimization problem, which can be solved through an exhaustive search method (ESM).
However, this method requires to search over all ${Q}$ combinations of the rotation angles and its complexity can be prohibitive for large $L$ and/or ${N_{\rm MA}}$ values.
To reduce the computational complexity, we propose an alternative method that solves problem \eqref{Problem} sub-optimally but with much lower complexity compared to the ESM, given in the next section. 

\vspace{-0pt}
\section{Proposed Algorithm}
To solve problem \eqref{Problem}, we apply the AMCMC-based method with Metropolized independence sampler (MIS) \cite{liu2001monte}.
We first adopt (\ref{C}) to formulate an original probability density function (PDF), given by $\xi ({{\boldsymbol{\epsilon}}})=\exp (  \frac{1}{{\tau}}{{{\hat{ C}}({\boldsymbol{ \varphi} }_{{\boldsymbol{\epsilon}}})}})/\Lambda$, where $\tau$ denotes a rate constant and $\Lambda$ denotes a normalizing factor to ensure $\sum_{{\boldsymbol{\epsilon}} \in {\mathcal{Z}}}\xi ({{\boldsymbol{\epsilon}}})=1$.
Thus, the problem in (\ref{Problem}) can be reformulate as
\vspace{-2pt}
\begin{equation}\label{Problem2}
	\vspace{-2pt}
	{\boldsymbol{\epsilon}}^*=\arg \mathop {\max }\limits_{{\boldsymbol{\epsilon}} \in {\mathcal{Z}} } ~ \xi ({{\boldsymbol{\epsilon}}}).
	\vspace{-1 pt}
\end{equation}
To solve the problem in (\ref{Problem2}), we consider the MIS for the AMCMC method, which is described as follows \cite{liu2001monte}: given any current indicator vector ${\boldsymbol{\epsilon}}_s\in\mathcal{Z}$ at the $s$-th (inner) iteration, we generate a new indicator vector ${\boldsymbol{\epsilon}}_{\rm new}\in\mathcal{Z}$ according to a proposal distribution $\Gamma ({\boldsymbol{\epsilon}};\bf p)$.
Then, based on an accepting probability $p_{\rm ac}({\boldsymbol{\epsilon}}_{\rm new},{\boldsymbol{\epsilon}}_s;{\bf p})= \min \big\{ {1,\frac{\xi ({\boldsymbol{\epsilon}}_{\rm new})}{\xi ({\boldsymbol{\epsilon}}_{s})} \frac{\Gamma ({\boldsymbol{\epsilon}}_{s};{\bf p})}{\Gamma ({\boldsymbol{\epsilon}}_{\rm new};{\bf p})}} \big\} $, the indicator vector sample for the next iteration will be ${\boldsymbol{\epsilon}}_{s+1}={\boldsymbol{\epsilon}}_{\rm new}$, if accepted, or ${\boldsymbol{\epsilon}}_{s+1}={\boldsymbol{\epsilon}}_{s}$, otherwise.
By repeating the above after $N_s$ (inner) iterations, i.e., $s=0,1,\cdots,N_s-1$, we obtain a set of $(1 + N_s)$ indicator vectors including the initial indicator vector ${\boldsymbol{\epsilon}}_0$ for $s=0$, i.e., $\{{\boldsymbol{\epsilon}}_0,{\boldsymbol{\epsilon}}_1,\cdots,{\boldsymbol{\epsilon}}_{N_s}\}$.

For the proposal distribution, we adopt the product of Bernoulli distributions, which is given by 
	\vspace{-4 pt}
\begin{equation}\label{q}\vspace{-2 pt}
	\Gamma ({\boldsymbol{\epsilon}};{\bf p})=\dfrac{1}{\Lambda '}\prod\limits_{l = 1}^{{L}} {p_{l}^{\left[{\boldsymbol{\epsilon}}\right]_{l}}\left(1-p_{l}\right)^{1-\left[{\boldsymbol{\epsilon}}\right]_{l}}} ,
\end{equation}
where ${\bf p}\in\mathbb{R}^{L\times 1}$ with its entry, $p_{l}, l\in\mathcal{L}$, denoting the probability of the $l$-th candidate rotation angle, ${\tilde \varphi}_l$, to be selected, and $\Lambda '$ denotes a normalizing factor to ensure $\sum_{{\boldsymbol{\epsilon}} \in {\mathcal{Z}}}\Gamma ({{\boldsymbol{\epsilon}}})=1$.
Since $\Lambda $ and $\Lambda '$ are both canceled out in computing the accepting probability $p_{\rm ac}$, their values do not need to be computed.
To increase the similarity between $\xi ({{\boldsymbol{\epsilon}}})$ and $\Gamma ({\boldsymbol{\epsilon}};{\bf p})$, we update the probability entry $p_{l}$ to update $\Gamma ({\boldsymbol{\epsilon}};{\bf p})$ by minimizing the Kullback-Leibler divergence between $\xi ({{\boldsymbol{\epsilon}}})$ and $\Gamma ({\boldsymbol{\epsilon}};{\bf p})$.
The recursive update equation is given by  \cite{liu2001monte}
\vspace{-2 pt}
\begin{equation}\label{p_l}
		\vspace{-2 pt}
	p_{l}^{(t+1)}=p_{l}^{(t)}+\alpha^{(t+1)}\bigg(\dfrac{1}{N_{s}}\sum\limits_{s=1}^{N_{s}} {\big[{\boldsymbol{\epsilon}}_s\big]_{l}}-p_{l}^{(t)}  \bigg),
\end{equation}
where superscript $(\cdot)^{(t)}$ denotes the $t$-th (outer) iteration for updating the probability, and $\alpha^{(t)}=\frac{1}{N_{s}+t}$ is a sequence of decreasing step sizes that satisfies $\sum_{t=0}^{\infty}\alpha^{(t)}=\infty$ and $\sum_{t=0}^{\infty}({\alpha^{(t)}})^2<\infty$.
The AMCMC-based algorithm for solving problem (\ref{Problem2}) is summarized in Algorithm 1, where $T$ denotes the maximum number of (outer) iterations. 
The complexity order of the above algorithm is $\mathcal{O}(N_s T M^2\tilde K\Upsilon)$ with $\tilde{K}={\rm max}(K_1,K_2,\cdots,K_\Upsilon)$, which is much lower than that of the ESM, i.e., $\mathcal{O}( QM^2\tilde K\Upsilon)$, as in general $Q\gg N_sT$.
\begin{algorithm}[!t]
	\renewcommand{\thealgorithm}{1:}
	\caption{The AMCMC-based algorithm for solving problem (\ref{Problem2})}\label{alg:alg1}
	\begin{algorithmic}[1]
		\STATE \textbf{Initialization:} Set ${\boldsymbol{\epsilon}}_0=[{\bf 1}_{N_{\rm MA}}^T,{\bf 0}_{L-N_{\rm MA}}^T]^T$, ${\boldsymbol{\epsilon}}^*={\boldsymbol{\epsilon}}_{0}$ and $p_{l}^{(0)}=\frac{1}{2}, \forall l\in\mathcal{L}$.
		\STATE {\textbf{for}} $t=1:1:T$ {\textbf{do}} 
		\STATE \hspace{0.2cm} {\textbf{for}} $s=0:1:N_s-1$ {\textbf{do}} 
		\STATE \hspace{0.5cm} Generate ${\boldsymbol{\epsilon}}_{\rm new}$ based on $\Gamma ({\boldsymbol{\epsilon}};{\bf p}^{(t-1)})$.
		\STATE \hspace{0.5cm} Generate $u$ based on  $U\left[ 0,1\right]$.
		\STATE \hspace{0.9cm} {\textbf{if}} $u<p_{\rm ac}({\boldsymbol{\epsilon}}_{\rm new},{\boldsymbol{\epsilon}}_s;{\bf p}^{(t-1)})$ {\textbf{then}}
		\STATE \hspace{1.2cm} ${\boldsymbol{\epsilon}}_{s+1}\leftarrow {\boldsymbol{\epsilon}}_{\rm new}$.
		\STATE \hspace{0.9cm} {\textbf{else}} 
		\STATE \hspace{1.2cm} ${\boldsymbol{\epsilon}}_{s+1}\leftarrow{\boldsymbol{\epsilon}}_{s}$.
		\STATE \hspace{0.9cm} {\textbf{end if}} 
		\STATE \hspace{0.9cm} {\textbf{if}} $\xi ({{\boldsymbol{\epsilon}}}_{s+1})>\xi ({{\boldsymbol{\epsilon}}}^*)$ {\textbf{then}}
		\STATE \hspace{1.2cm} ${\boldsymbol{\epsilon}}^*\leftarrow{\boldsymbol{\epsilon}}_{s+1}$.
		\STATE \hspace{0.9cm} {\textbf{end if}} 
		\STATE \hspace{0.2cm} {\textbf{end for}}
		\STATE \hspace{0.2cm} Update ${\bf p}^{(t)}$ via (\ref{p_l}).
		\STATE \hspace{0.2cm} ${\boldsymbol{\epsilon}}_0\leftarrow{\boldsymbol{\epsilon}}^*$.
		\STATE {\textbf{end for}}
		\STATE \textbf{return}  ${\boldsymbol{\epsilon}}^*$.
	\end{algorithmic}
	\label{alg1}
\end{algorithm}

\vspace{-2 pt}
\section{Simulation Results}
\vspace{-0 pt}
In this section, we present numerical results to validate the performance of the proposed HFMA-BS scheme with the AMCMC-based algorithm.
Unless otherwise stated, the simulation settings are as follows.
We set ${{\phi _{3{\rm{dB}}}}}={65^ \circ }$, $G_m=0~{\rm dBi}$, ${G_s}=25~{\rm dBi}$,  $\beta_0=-40~\rm dB$, $P_0=1~\rm mW$, $\sigma ^2\!=\!-80~\rm dBm$, $\lambda_c\!=\!0.125~\rm m$, $D_{\rm min}\!=\! \frac{\lambda_c}{2}M_{{\rm MA},h}$, $r_1=r_2=1~\rm m$, $h_{1}=9~\rm m$, $h_2=10~\rm m$, $N_{\text{MA}}=6$, $L=40$, $M_{{\rm MA}}=2\times8$, $M_{{\rm FA}}=8\times8$, $\Upsilon=100$, and   $N_s=20$, $T=10$ (for Algorithm 1).
The fixed rotation angles of three FPA arrays are set as $\frac{\pi}{2}$, $\frac{7\pi}{6}$ and $\frac{11\pi}{6}$, respectively.
For the user distribution, we set the entire cell area $\mathcal{A}_0$ as a 2D disk centered at origin $\bf o$ of the CCS, with a radius of $R_0=100\ \rm m$.
Within $\mathcal{A}_0$, there are $W=3$ hotspot areas, i.e., $\mathcal{A}_1$, $\mathcal{A}_2$ and  $\mathcal{A}_3$.
Each hotspot area is set as a 2D disk with a radius $R_w$, centered at a distance $D_w$ and an azimuth angle $\psi_w$ with respect to the origin $\bf o$,  denoted by $\mathcal{A}_w={\bf b}(\psi_w,D_w,R_w), w=1,2,3$.
We set $\mathcal{A}_1={\bf b}(\frac{\pi}{4},50{\ \rm m},10{\ \rm m})$, $\mathcal{A}_2={\bf b}(\frac{7\pi}{6},60{\ \rm m},15{\ \rm m})$ and $\mathcal{A}_3={\bf b}(\frac{7\pi}{4},70{\ \rm m},20{\ \rm m})$.
The average number of users, denoted by $\bar{K}$, is set as $300$, with $\bar K=\sum_{w=0}^{3}\mu_{w}\pi R_w^2$. 
In addition, the average numbers of hotpots users, denoted by $\bar K_w,w=1,2,3$, follow the ratio of $1:2:3$ with their sum set equal to $\sum_{w=1}^3{\bar K_w}=\frac{\bar K}{2}$.
We consider the following benchmark schemes: {\footnote{The default system parameters have been set to ensure that the variables in each benchmark scheme are integers.}}
\begin{itemize}
	\item Scheme 1:
	This scheme sets the rotation angles of 6DMA surfaces as the $B_w$ candidate rotation angles closest to each of the $W$ $\psi_w$'s, with $B_w=\frac{N_{\rm MA}}{W},w=1,2,3$.
\end{itemize}
\begin{itemize}
	\item Scheme 2:
	Based on Scheme 1, this scheme further sets $B_w$'s according to $B_1:B_2:B_3={\bar K_1}:{\bar K_2}:{\bar K_3}$.
\end{itemize}
\begin{itemize}
	\item Scheme 3: This scheme only considers three FPA arrays, each equipped with ${\frac{M}{3}} $ antennas (thus, the total number of antennas is the same as in the HFMA-based schemes).
\end{itemize}

\begin{figure*}[t]
	%\setcaptionwidth{0.3\textwidth}
	\centering
	\begin{minipage}[t]{0.33\textwidth}
		\centering
		\includegraphics[width=2.2in]{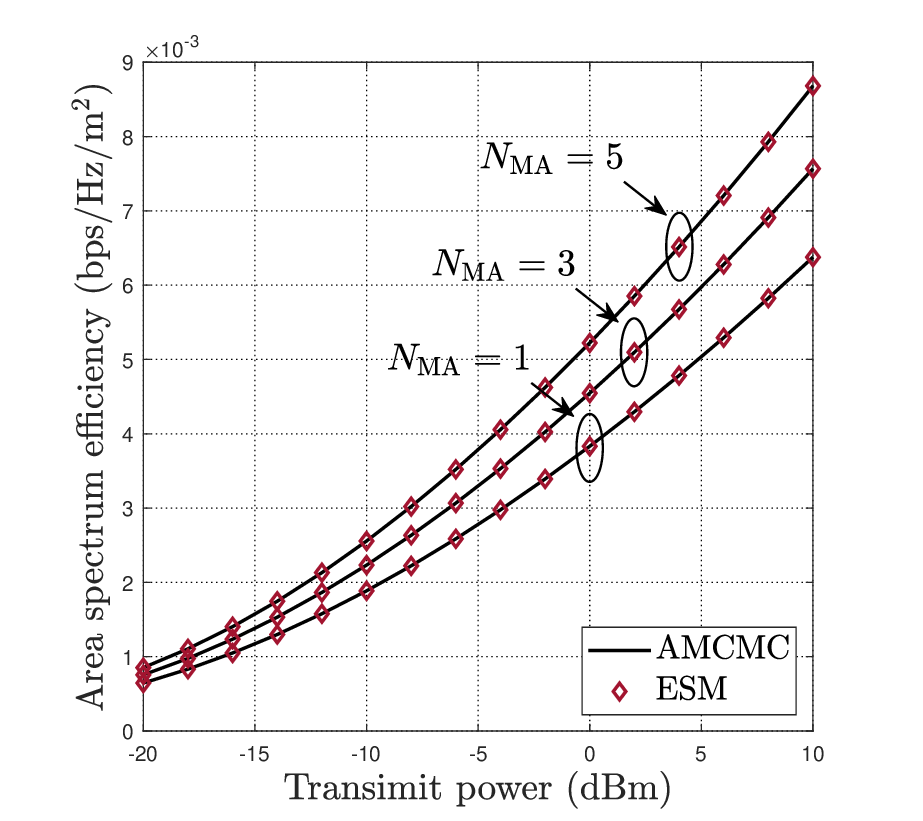}
	\end{minipage}%
	\begin{minipage}[t]{0.33\textwidth}
		\centering
		\includegraphics[width=2.2in]{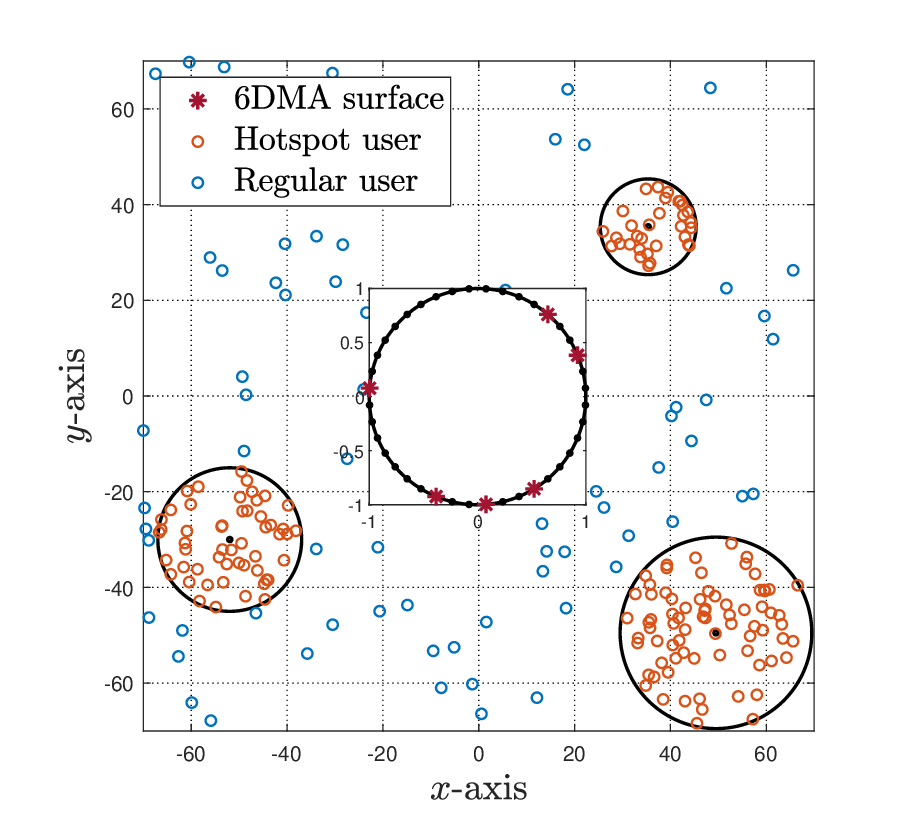}
	\end{minipage}%
	\begin{minipage}[t]{0.33\textwidth}
		\centering
		\includegraphics[width=2.2in]{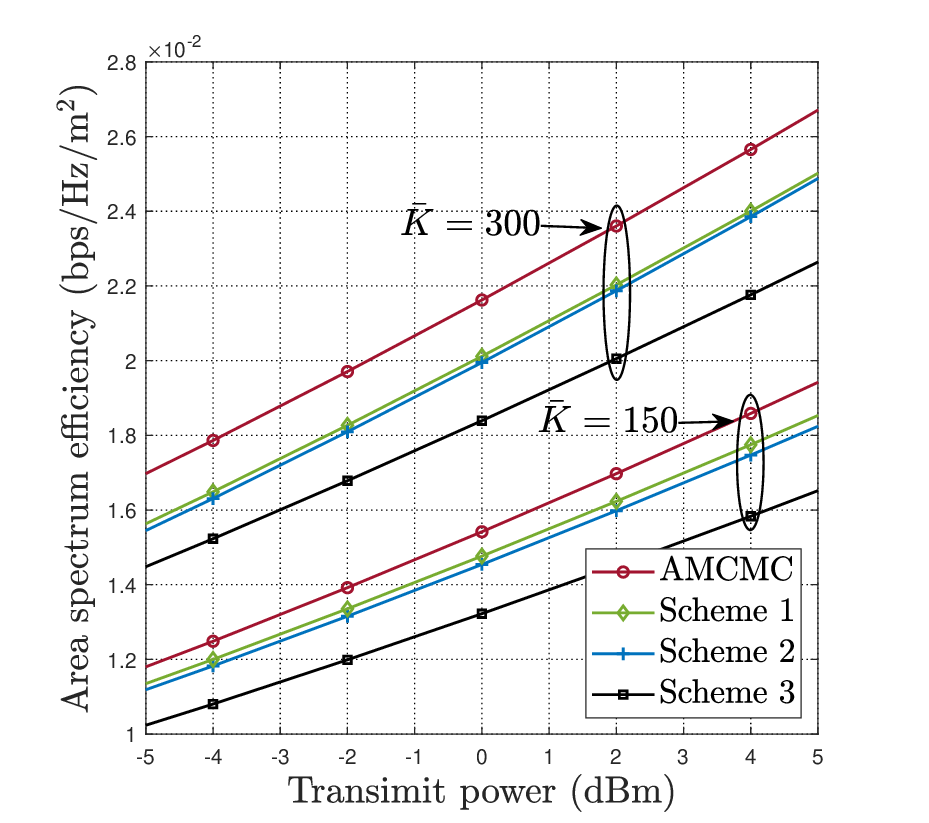}
	\end{minipage}\\[-10pt]
	\begin{minipage}[t]{0.33\textwidth}
		\vspace{-2 mm}
		\caption{ASE versus transmit power per user with  $L=20$, $M_{{\rm MA}}=2\!\times\!2$, $M_{{\rm FA}}=4\!\times\!4$ and $\bar K=75$.}
		\label{f1_vsESM}
	\end{minipage}%
	\hspace{2 mm}
	\begin{minipage}[t]{0.3\textwidth}
		\vspace{-2 mm}
		\caption{Optimized rotations of 6DMA surfaces by the AMCMC-based algorithm.}
		\label{f3_positions}
	\end{minipage}
	\hspace{2 mm}
	\begin{minipage}[t]{0.3\textwidth}
		\vspace{-2 mm}
		\caption{ASE versus transmit power per user for different schemes with different  $\bar{K}$.}
		\label{f4_benchmark}
	\end{minipage}
\end{figure*}

Fig. \ref{f1_vsESM} illustrates the area spectrum efficiency (ASE), defined as ${{\hat C}({  {\boldsymbol{ \varphi} }})}/({\pi R_0^2})$, versus the transmit power, $P_0$.
It is observed that the AMCMC-based algorithm achieves very close performance to the ESM, for different values of $N_{\text{MA}}$ under $L=20$. 
Since the AMCMC-based algorithm has much lower complexity than the ESM, we adopt the former for subsequent simulations with larger values of $N_{\text{MA}}=6$ and $L=40$.
\begin{comment}
\begin{figure}[!t]
	\centering
	\includegraphics[width=2.8in]{f2.eps}
	\vspace{-10pt}
	\caption{ASE versus the transmit power per user with  $L=20$, $M_{{\rm MA}}=2\times2$, $M_{{\rm FA}}=4\times4$ and $\bar K=75$.}
	\vspace{-5pt}
	\label{f1_vsESM}
\end{figure}

\begin{figure}[!t]
	\centering
	\includegraphics[width=2.8in]{f3.eps}
	\vspace{-10pt}
	\caption{The optimized rotations of 6DMA surfaces using AMCMC algorithm.}
	\label{f3_positions}
\end{figure}
\begin{figure}[!t]
	\centering
	\includegraphics[width=2.8in]{f4.eps}
	\vspace{-10pt}
	\caption{ASE versus the transmit power per user.}
	\label{f4_benchmark}
\end{figure}
\end{comment}

Fig. \ref{f3_positions} shows a visual representation of the optimal rotations of 6DMA surfaces obtained by the AMCMC-based algorithm.
It can be observed that their rotations strike a balance between serving hotspot users and the users positioned in the angles between the adjacent FPA arrays to maximize the network capacity.  
Interestingly, it is also observed that under the non-uniform user spatial distribution, rotating 6DMA surfaces towards hotspot areas only (i.e., Schemes 1 and 2) may not be capacity-optimal for the HFMA-BS.

In Fig. \ref{f4_benchmark}, we compare the ASE obtained by the AMCMC-based algorithm with three benchmark schemes.
It is observed that the proposed scheme achieves significantly improved ASE
over all benchmark schemes. 
Furthermore, their performance gap in the case with $\bar K=300$ is larger than that in the case with $\bar K=150$.
This is because, when the user density increases, rotating 6DMA surfaces to exploit their array gain and spatial multiplexing gain becomes more crucial to maximizing the network capacity, under the setting with non-uniform user spatial distribution.

\vspace{-6pt}
\section{Conclusion}
\vspace{-1pt}
This letter proposed a new and cost-efficient HFMA-BS architecture consisting of both conventional FPA arrays and position/rotation-adjustable 6DMA surfaces. 
We aimed to maximize the average network capacity via optimizing the rotation angles of all 6DMA surfaces along a circular track.
To solve this combinatorial optimization problem, we proposed the AMCMC-based algorithm with much lower complexity compared to the ESM.
Numerical results verified that the HFMA-BS with optimized 6DMA rotations can drastically improve the network capacity compared to the conventional BS with  FPAs only or the HFMA-BS with heuristic 6DMA rotations.
Moreover, the performance gain was shown to be more appealing when the average user density increases under the non-uniform user spatial distribution.
\vspace{-2pt}
\bibliographystyle{IEEEtran}
\bibliography{123}
\end{document}